%% file: main.tex
\documentclass{llncs}
\usepackage{amsfonts, amsmath}
\usepackage{graphicx,color}
\usepackage{algorithm}
\usepackage{algorithmic}
\usepackage{tikz}
\usepackage{booktabs}
\usepackage{enumitem}
\usepackage[utf8]{inputenc}
\usepackage{setspace}
\usepackage{multirow}
\usepackage{xspace}  
\usepackage{xargs}                      
\usepackage{wrapfig}
\usepackage{enumitem}

\input macros

\newcommand{\approach}{CPSDebug\xspace}

\title{Automatic Failure Explanation in CPS Models}
\author{}
\institute{}
\author{Ezio Bartocci\inst{1} \and Niveditha Manjunath\inst{1,2} \and Leonardo Mariani\inst{3} \and Cristinel Mateis\inst{2} \and Dejan Ni\v{c}kovi\'{c}\inst{2}}
\institute{Vienna University of Technology \and AIT Austrian Institute of Technology \and University of Milano-Bicocca}

\begin{document}
\maketitle

\begin{abstract}

Debugging Cyber-Physical System (CPS) models can be extremely complex. 
Indeed, only the detection of a failure is insufficient to know how to correct a faulty 
model. Faults can propagate in time and in space producing observable 
misbehaviours in locations completely different from the location of the fault. 
Understanding the reason of an observed failure is typically a challenging and 
laborious task left to the experience and domain knowledge of the designer.

In this paper, we propose \approach, a novel approach that by combining
testing, specification mining, and failure analysis, can automatically explain failures 
in Simulink/Stateflow models.
We evaluate \approach on two case studies, involving two use scenarios and several 
classes of faults, demonstrating the potential value of our approach.

\end{abstract}

\input intro

\input background
\input running
\input localization
\input experimental

\input related
\input conclusions

\bibliographystyle{plain}
\bibliography{main}

\end{document}

%% file: macros.tex
\let\phi\varphi

\newcommand\f\varphi
\newcommand\g\psi

\newcommand{\R}{\mathbb{R}}

\newcommand{\T}{\mathbb{T}}

\newcommand\tees{\Theta}

\DeclareMathOperator\eventually{\lozenge}
\DeclareMathOperator\always{\Box}

\DeclareMathOperator\until{\mathcal{U}}
\DeclareMathOperator\since{\mathcal{S}}
\newcommand\true{\top}

\newcommand\ares{R}
\newcommand\esses{S}

\usepackage[colorinlistoftodos,prependcaption,textsize=tiny]{todonotes}
\newcommandx{\unsure}[2][1=]{\todo[linecolor=red,backgroundcolor=red!25,bordercolor=red,#1]{#2}}
\newcommandx{\change}[2][1=]{\todo[linecolor=blue,backgroundcolor=blue!25,bordercolor=blue,#1]{#2}}
\newcommandx{\info}[2][1=]{\todo[linecolor=OliveGreen,backgroundcolor=OliveGreen!25,bordercolor=OliveGreen,#1]{#2}}
\newcommandx{\improvement}[2][1=]{\todo[linecolor=Plum,backgroundcolor=Plum!25,bordercolor=Plum,#1]{#2}}
\newcommandx{\thiswillnotshow}[2][1=]{\todo[disable,#1]{#2}}

%% file: intro.tex

\section{Introduction}
\label{sec:intro}

Cyber-Physical Systems (CPS) combine computational and physical entities that interact with sophisticated and unpredictable environments via sensors and actuators. To cost-efficiently study their behavior, engineers typically apply model-based development methodologies, which combine modeling and simulation activities with prototyping.  The successful development of CPS is thus strongly dependent on the quality and correctness of their models.

CPS models can be extremely complex: they may include hundreds of variables, signals, look-up tables and components, combining continuous and discrete dynamics. Verification and testing activities are thus of critical importance to early detect problems in the models~\cite{BartocciFMN18,DeshmukhJMP18,staliro,LiuLNB17,LiuLNBB16}, before they propagate to the actual CPS.  Discovering faults is however only part of the problem. Due to their complexity, debugging the CPS models by identifying the causes of failures can be as challenging as identifying the problems themselves~\cite{Lee08a}. 

CPS functionalities are often modelled using the MathWorks\texttrademark\; Simulink toolset.A well-established approach to find bugs in Simulink/Stateflow models is using \emph{falsification-based testing}~\cite{staliro,Nghiem+Others/2010/Monte,SankaranarayananF12}.  This approach is based  on quantifying (by monitoring~\cite{BartocciDDFMNS18}) how much a simulated trace of CPS behavior is close to violate a requirement expressed in a formal specification language, such as Signal Temporal Logic (STL)~\cite{mn04}. This measure enables the systematic exploration of  the input space searching  for the first input sequence responsible for a violation.  However, this method does  not provide any suitable information  about which component should be inspected to resolve the violation.  Trace diagnostics~\cite{FerrereMalerNickovic15} identifies (small)  segments of the observable model behavior that are sufficient to imply the violation of the formula, thus providing  a failure explanation at the input/output model interface. However, this is a black-box technique that does not  attempt to open the model and explain the failure in terms of its internal signals and components.   Other approaches are based on \emph{fault-localization}~\cite{BartocciFMN18,DeshmukhJMP18,LiuLNBB16,LiuLNBB16b,LiuLNB17}, a statistical technique measuring the code coverage in the failed 
and successful tests.  This method provides a limited explanation that does not  often help the engineers to understand if the selected code  is really faulty and how the fault has propagated across the components resulting on actual failure.

In this paper, we advance the knowledge in failure analysis of CPS models by presenting \approach, a technique that originally combines testing, specification mining, and failure analysis. \approach first exercises the CPS model under analysis by running the available test cases, while discriminating passing and failing executions using requirements formalized as a set of STL formulas.  While running the test cases, \approach  records the internal behavior of the CPS model, that is, it records the values of all the internal system variables at every timestamp. It then uses the values collected from passing test cases to infer properties about the variables and components involved in the computations. These properties capture how the model behaves when the system runs correctly.

\approach checks the mined properties against the traces collected for the failed test cases to discover the internal variables, and corresponding components, that are responsible for the violation of the requirements. Finally, failure evidence is analyzed using trace diagnostics~\cite{FerrereMalerNickovic15} and clustering~\cite{HastieTF09}  to produce a time-ordered sequence of snapshots that show where the anomalous variables  values originated and how they propagated within the system.

\approach thus overcomes the limitation of state of the art approaches that do not guide engineers in their analysis, but only indicate the inputs or the code locations that might be responsible for the failure. On the contrary, the sequence of snapshots returned by \approach provides a step by step illustration of the failure with explicit indication of the faulty behaviors. Our evaluation involved with three classes of faults, two actual CPS models, and feedback from industry engineers confirmed that  the output produced by \approach can be indeed valuable to ease  the failure analysis and debugging process.

The rest of the paper is organized as follows.  We provide background information in Section~\ref{sec:background} and we describe the case study in Section~\ref{sec:caseStudy}.  In Section~\ref{sec:failure_explanation} we present our approach for failure explanation while in Section~\ref{sec:experimental}  we provide the empirical evaluation. We discuss the related work in Section~\ref{sec:related}  and we draw our conclusions in Section~\ref{sec:conclusions}.

%% file: background.tex
\section{Background}
\label{sec:background}

\subsection{Signals and Signal Temporal Logic}

We define $\esses = \{s_1,\hdots,s_n\}$ to be a set of signal variables.
A \emph{signal} or \emph{trace} $w$ is a function $\T \to \R^n$, where $\T$ is the time domain in the form  of $[0,d]\subset \R$.
We can also see a multi-dimensional signal $w$ as a vector of real-valued uni-dimensional signals $w_i : \T \to \R$ associated to variables $s_i$ for $i=1,\hdots,n$. 
We assume that every signal $w_i$ is piecewise-linear. Given two signals $u : \T \to \R^l$ and $v : \T \to \R^m$, 
we define their parallel composition $u \| v : \T \to \R^{l+m}$ in the expected way.
Given a signal $w : \T \to \R^n$ defined over the set of variables $\esses$ and a subset of variables $\ares \subseteq \esses$, we denote by $w_{\ares}$ the projection of 
$w$ to $\ares$, where $w_{\ares} = \|_{s_i \in \ares} w_i$.

Let $\tees$ be a set of terms of the form $f(\ares)$ where $\ares \subseteq \esses$ are subsets of variables and $f : \R^{|\ares|} \to \R$ are interpreted functions.
The syntax of STL is defined by the grammar:
\begin{align*}
\f ::= \true \mid f(\ares) > 0 \mid \lnot \f \mid \f_1 \lor \f_2 \mid \f_1 \until_I \f_2 \,,
\end{align*}
where $f(R)$ are terms in $\tees$ and $I$ are real intervals with bounds in $\mathbb Q_{\ge 0} \cup \{\infty\}$.
As customary we use the shorthands $\eventually_I \f \equiv \true \until_I \f$ for \emph{eventually}, $\always_I \f \equiv \lnot \eventually_I \lnot \f$ for \emph{always}, 
$\uparrow \f \equiv \f \wedge \true \since \neg \f$ for \emph{rising edge} and $\uparrow \f \equiv \neg \f \wedge \true \since \f$ for \emph{falling edge}\footnote{We omit the timing modality $I$ when $I=[0,\infty)$.}. We interpret STL with its classical semantics defined in~\cite{DBLP:journals/sttt/MalerN13}.

\subsection{Daikon}
Daikon is a template-based property inference technique that, starting from a set of variables and a set of observations, can infer a set of properties that are likely to hold for the input variables. More formally, given a set of variables $V=V_1,\ldots ,V_n$ defined in the domains $D_1,\ldots D_n$, an observation for these variables is a tuple $\overline{v}=(v_1, \ldots, v_n)$, with $v_i \in D_i$. 

Given a set of variables $V$ and multiple observations $\overline{v}_1 \ldots \overline{v}_m$ for these same variables, Daikon is a function $D(V,\overline{v}_1 \ldots \overline{v}_m)$ that returns a set of properties $\{p_1, \ldots p_k\}$, such that $\overline{v_i} \models p_j \forall i,j$, that is, all the observations satisfy the inferred properties. For example, considering two variables $x$ and $y$ and considering the observations $(1,3)$, $(2,2)$, $(4,0)$ for the tuple $(x,y)$, Daikon can infer properties such as $x>0$, $x+y=4$, and $y\geq0$.

The inference of the properties is driven by a set of template operators that Daikon instantiates over the input variables and checks against the input data. 
Since template-based inference can generate redundant and implied properties, Daikon automatically detects them and reports the relevant
properties only.
Finally, to guarantee that the inferred properties are relevant, Daikon computes the probability that the inferred property holds by chance for all the properties. Only if the property is statistically significant with a probability higher than $0.99$ the property is assumed to be reliable and it is reported in the output, otherwise it is suppressed.

In our approach, we use Daikon to automatically generate fine-grained properties that capture the behavior of the individual components and individual signals in the model under analysis. These properties can be used to precisely detect misbehaviours and their propagation.

%% file: running.tex
\section{Case Study} \label{sec:caseStudy}

We now introduce a case study that we use as a running example to illustrate 
our approach step by step.  The case study is an aircraft elevator control system, 
introduced in~\cite{mosterman-fdir}, to illustrate model-based development of a 
fault detection, isolation and recovery (FDIR) application for a redundant actuator 
control system. 

\begin{figure}
\centering
\scalebox{0.65}{ 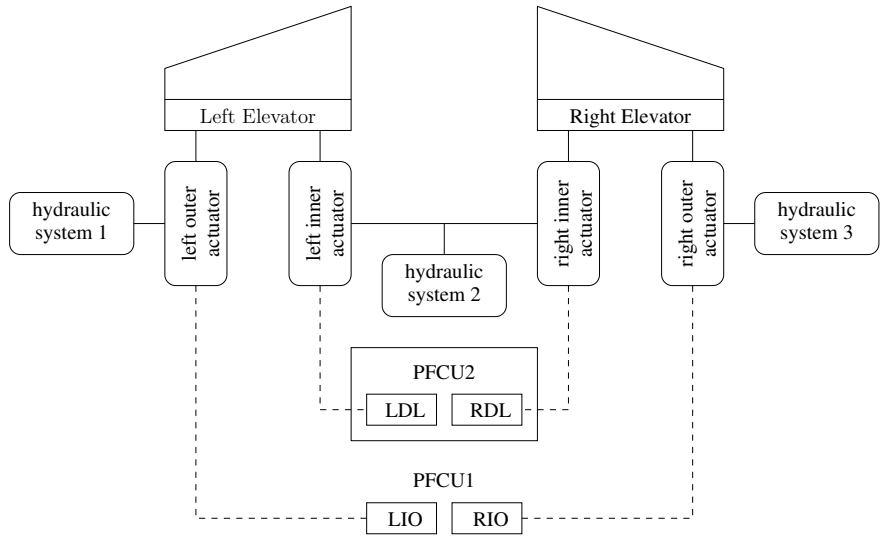 }
\caption{Aircraft elevator control system~\cite{mosterman-fdir}.}
\label{fig:aircraft}
\vspace{-4ex}
\end{figure}

Figure~\ref{fig:aircraft}  shows the architecture of the aircraft elevator control system 
with redundancy, with one elevator on the left and one on the right side. Each elevator 
is equipped with two hydraulic actuators. Both actuators can position the elevator, but 
only one shall be active at any point in time. There are three different hydraulic circuits 
that drive the four actuators. The left (LIO) and right (RIO) outer actuators 
are controlled by a Primary Flight Control Unit (PFCU) with a sophisticated input/output 
control law.  If a failure happens, a less sophisticated Direct-Link (DL) control law with 
reduced functionality takes over to handle the left (LDL) and right (RDL) inner actuators. 
The system uses  state machines to coordinate the redundancy and assure its continual 
fail-operational activity. 

This model has one input variable, the input pilot command, and two output variables, 
the position of the left and right actuators, as measured by the sensors. 
This is a complex model that could be extremely hard to analyze in case of failure. In fact, 
the model has $426$ signals, from which $361$ are internal variables that are instrumented 
($279$ real-valued, $62$ Boolean and $20$ enumerated - state machine - variables) and 
any of them, or even a combination of them, might be responsible for an observed failure.

The model comes with a failure injection mechanism, which allows to dynamically insert 
failures that represent hardware/ageing problems into different components of the system 
during its simulation. This mechanism allows insertion of (1) low pressure failures for each 
of the three hydraulic systems, and (2) failures of sensor position components in each of 
the four actuators. Due to the use of the redundancy in the design of the control system, a 
single failure is not sufficient to alter its intended behavior. In some cases even two failures 
are not sufficient to produce faulty behaviors. For instance, the control  system is able to 
correctly function when both a left and a right sensor position components simultaneously 
fail. This challenges the understanding of failures because there are multiple 
causes that must be identified to explain a single failure. 

To present our approach we consider the analysis of a system failure caused by the activation 
of two failures: the sensor measuring the left outer actuator position 
failing at time $2$ and the sensor measuring the left inner actuator position failing at time $4$. 
To collect evidence of how the system behaves, we executed the Simulink model with $150$ 
test cases with different pilot commands and collected the input-output behavior both with 
and without the failures. 

When the system behaves correctly, the intended position of the aircraft required by the pilot 
must be achieved within a predetermined time limit and with a certain accuracy. This can be 
captured with several requirements. One of them says that whenever pilot command $\emph{cmd}$ 
goes above a threshold $m$, the actuator position measured by the sensor must stabilize 
(become at most $n$ units away from the command signal) within 
$T+t$ time units. This requirement is formalized in STL with the following 
specification:
$$
\varphi \equiv \always (\uparrow(\emph{cmd} \geq m) \rightarrow \eventually_{[0,T]} \always_{[0,t]} (|\emph{cmd} - \emph{pos}| \leq n)).
$$
\begin{figure}[h]
\begin{center}
  \begin{minipage}[b]{0.48\columnwidth}
    \includegraphics[width=\columnwidth]{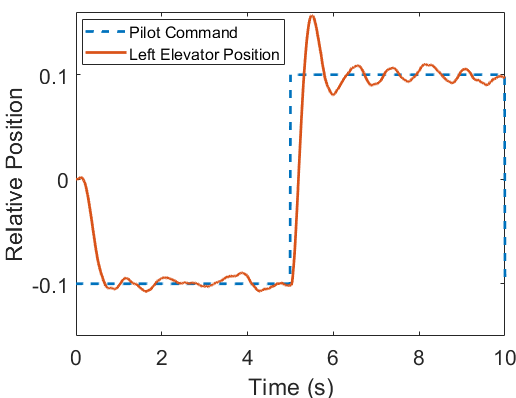}
    \caption{Expected behavior of the aircraft control system.}
    \label{fig:expected}
  \end{minipage}
  \hspace{0.1cm}
  \begin{minipage}[b]{0.48\columnwidth}
    \includegraphics[width=\columnwidth]{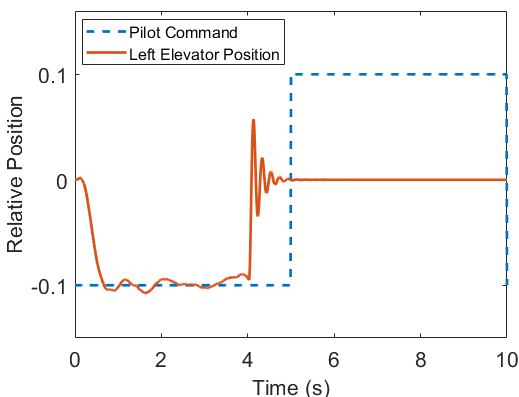}
    \caption{Failure of the aircraft control system.}
    \label{fig:faulty}
  \end{minipage}  
  \end{center}
\vspace{-20pt}
\end{figure}
Figures~\ref{fig:expected} and~\ref{fig:faulty} shows the correct and faulty behavior of the system. The control system clearly stops following the reference signal after $4$ seconds. The failure observed on the input/output interface of the model does not give any indication within the model on the reason leading to the property violation. 
In the next section, we present how our failure explanation technique can address this case producing a valuable output to engineers.

%% file: aircraft.pdf_t
\begin{picture}(0,0)%
\includegraphics{aircraft.pdf}%
\end{picture}%
\setlength{\unitlength}{3947sp}%
\begingroup\makeatletter\ifx\SetFigFont\undefined%
\gdef\SetFigFont#1#2#3#4#5{%
  \reset@font\fontsize{#1}{#2pt}%
  \fontfamily{#3}\fontseries{#4}\fontshape{#5}%
  \selectfont}%
\fi\endgroup%
\begin{picture}(8424,5124)(589,-6673)
\put(3001,-2686){\makebox(0,0)[b]{\smash{{\SetFigFont{12}{14.4}{\rmdefault}{\mddefault}{\updefault}{\color[rgb]{0,0,0}Left Elevator}%
}}}}
\end{picture}%

%% file: localization.tex
\vspace{-10pt}
\section{Failure Explanation}
\label{sec:failure_explanation}

\begin{figure}[t]
\centering
\scalebox{0.67}{ 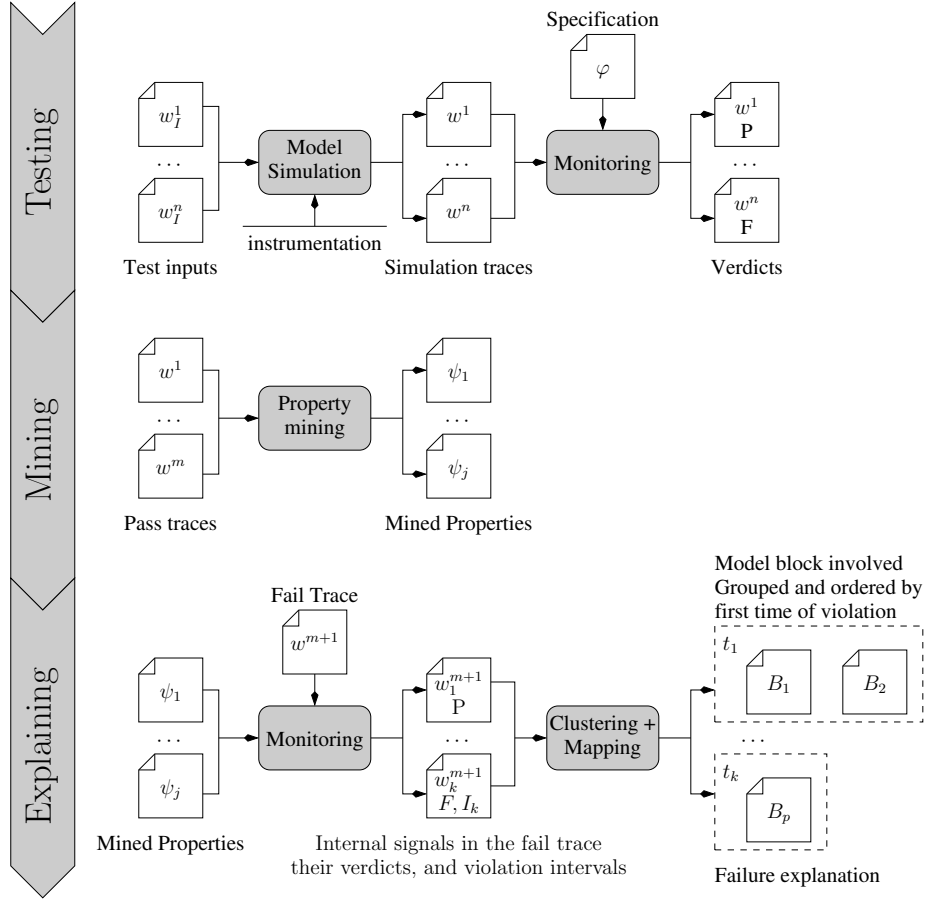 }
\caption{Overview of the failure explanation procedure.}
\label{fig:overview}
\vspace{-10pt}
\end{figure}

In this section we explain how \approach works with help
of the case study introduced in Section~\ref{sec:caseStudy}.
Figure~\ref{fig:overview} illustrates the main steps of the workflow.
Briefly, the workflow starts from a target CPS model and a test suite with some passing and failing test cases, and produces a failure explanation for each failing test case. The workflow consists of three sequential phases:

\begin{itemize}
\setlength\itemsep{0.4em}
\item[(i)] 
Testing - simulating the instrumented CPS model with available test cases to collect information about its behavior, both for passing and failing executions,
\item[(ii)] 
Mining - mining properties from the traces produced by the passing test cases; intuitively these properties capture the expected behavior of the model, 
\item[(iii)] 
Explaining - using the mined properties to analyze the traces produced by failures and generate failure explanations, which include information about the root events responsible for the failure and their propagation.
\end{itemize}

\vspace{-20pt}

\subsection{Testing}

\approach starts by instrumenting the CPS model. This is an important pre-processing step that is done before testing the model and that allows to 
log the internal signals in the model. The instrumentation is inductively defined on the hierarchical structure of the 
Simulink/Stateflow model and is done in a bottom-up fashion. For every signal variable having the real, Boolean or enumeration type, \approach assigns a unique name to it 
and makes the simulation engine to log its values. \approach also instruments the look-up tables and state machines in the model. It associates a dedicated variable to each look-up table. The variable is used to produce a simulation trace that records the unique cell index that is exercised by the input at every point in time. Similarly, \approach 
associates two dedicated variables per state-machine, one recording the transitions taken and one recording the locations visited during the simulation. 
We denote by $V$ the set of all instrumented model variables.

The first step of the testing phase, namely \emph{Model Simulation}, runs the available \emph{test cases} $\{w_I^k | 1 \leq k \leq n\}$ against the \emph{instrumented} version of the simulation model under analysis. The number of available test cases may vary case by case, for instance in our case study the test suite included  $n=150$ tests.

The result of the model simulation consists of one simulation trace $w^k$ for
each test case $w_I^k$, $1 \leq k \leq n$. The trace $w^k$ stores the sequence 
of (simulation time, value) pairs $w^k_v$ for every instrumented variable 
$v \in V$ collected during simulation. 

To determine the nature of each trace, we transform the informal model \emph{specification}, which is typically provided in form of 
free text, into an STL formula $\phi$ that can be
automatically evaluated by a \emph{monitor}. 
In fact, \approach checks every trace $w^k$ against the STL formula $\phi$, $1 \leq k \leq n$ and labels the trace with a \emph{pass verdict} if $w^k$ {\em satisfies} $\phi$, or 
a \emph{fail verdict} otherwise. In our case study, we had $149$ traces labeled as passing and one failing trace.

\vspace{-10pt}
\subsection{Mining}

In the mining phase, \approach selects the traces labeled with a pass verdict and exploits them for \emph{property mining}. 

Prior to the property inference, \approach performs several intermediate steps that facilitate the mining task. 
First, \approach reduces the set of variables $V$ to its subset $\hat{V}$ of {\em significant} variables by using cross-correlation. Intuitively, the presence of two highly correlated variables implies that one variable adds little information to the other one, and thus the analysis may actually focus on one variable only. 
The approach cross-correlates all passing simulation traces and whenever the cross-correlation between the
simulation traces associated with variables $v_{1}$ and $v_{2}$ in $V$ is higher than $0.99$, \approach removes one of the two variables 
(and its associated traces) from further analysis. In our case study, $|V| = 361$ and $|\hat{V}| = 121$, resulting in a reduction of $240$ variables.
 
In the next step, \approach associates each variable $v \in \hat{V}$ to (1) its domain $D$ and (2) its parent block $B$. We denote by $V_{D,B} \subseteq \hat{V}$ the 
set $\{v_{1}, \ldots, v_{m} \}$ of variables with the domain $D$ associated to block $B$. \approach collects all observations $\overline{v}_1 \ldots \overline{v}_n$ from 
all samples in all traces associated with variables in $V_{D,B}$ and uses the Daikon function $D(V_{D,B}, \overline{v}_1 \ldots \overline{v}_n)$ to infer a set of properties 
$\{p_{1}, \ldots, p_{k}\}$ related to the block $B$ and the domain $D$. The property mining per model block and model domain allows to avoid (1) combinatorial 
explosion of learned properties and (2) learning properties between incompatible domains.

Finally, \approach collects all the learned properties from all the blocks and the domains, and translates them to an STL specification, where each Daikon property $p$ is 
transformed to an STL assertion of type $\always p$.

In our case study, Daikon returned $96$ behavioral properties involving
$121$ variables, hence \approach generated an STL property $\psi$ with $96$ 
temporal assertions, i.e., $\psi = [\psi_1 \, \psi_2 \, ... \, \psi_{96}]$.
Table~\ref{tab:invariants} shows two examples of behavioral properties from our
case study inferred by Daikon and translated to STL. The first property states that 
the Mode signal is always in the state $2$ (Passive) or $3$ (Standby), while the 
second property states that the left inner position failure is encoded the same than the left outer position failure. 
\vspace{-10pt}
\begin{table}
$$
\begin{array}{lcl}
\varphi_{1} & \equiv & \always(\emph{mode} \in \{2, 3\}) \\
\varphi_{2} & \equiv & \always(\emph{LI\_pos\_fail} == \emph{LO\_pos\_fail} ) \\
\end{array}
$$
\caption{Examples of properties learned by Daikon. Variables $\emph{mode}$, $\emph{LI\_pos\_fail}$ and $\emph{LO\_pos\_fail}$
denote internal signals Mode, Left Inner Position Failure and Left Outer Position Failure from the aircraft position control Simulink model.}
\label{tab:invariants}
\end{table}
\vspace{-40pt}

\subsection{Explaining}

This phase analyzes a trace $w$ collected from a failing execution and produces a failure explanation. The \emph{Monitoring} step analyzes the trace against the mined properties and returns the signals that violate the properties and the time intervals in which the properties are violated. \approach subsequently labels with $F$ ({\em fail}) the internal signals involved in the violated properties and with $P$ ({\em pass}) the remaining signals from the trace. 
To each fail-annotated signal, \approach also assigns the violation time
intervals of the corresponding violated properties returned by the
monitoring tool. \\

In our case study, the analysis of the left inner and the left outer sensor failure resulted in the violation of $17$ mined
properties involving $19$ internal signals.

For each internal signal there can be several fail-annotated signal instances,
each one with a different violation time interval.
\approach selects the instance that occurs first in time, ignoring all other instances. 
This is because, to reach the root cause of a failure, \approach has to focus on the events that cause observable misbehaviours first.

Table~\ref{tab:summary} summarizes the set of property-violating signals, the block they belong to, 
and the instant of time the signal has first violated a property for our case study. We can observe that the $17$ signals participating in 
the violation of at least one mined property belong to only
$5$ different Simulink blocks. 
In addition, we can see that all the violations naturally cluster around 
two time instants -- $2$ seconds and $4$ seconds. This suggests that \approach can effectively isolate in space and time a limited number of events likely responsible for the failure. 

\begin{wrapfigure}{L}{0.55\textwidth}
\vspace{-20pt}
\centering
\includegraphics[scale=0.45]{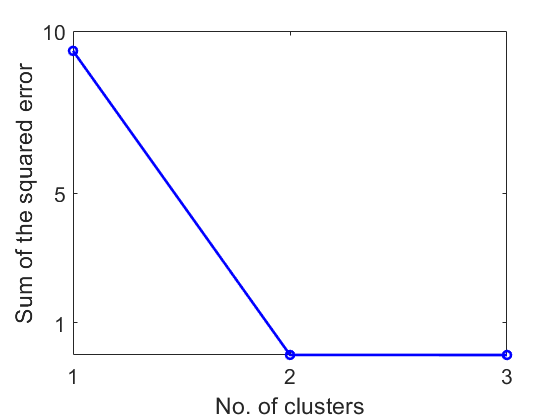}
\caption{Number of clusters versus the error. }
\label{fig:elbow}
\vspace{-20pt}
\end{wrapfigure}

The \emph{Clustering $\&$ Mapping} step then (i) clusters the resulting fail-annotated signal instances by their violation 
time intervals and (ii) maps them to the corresponding model blocks, i.e., to the model blocks that
have some of the fail-annotated signal instances as internal signals. 

\approach automatically derives the clusters by applying the elbow method
with the $k$-means clustering algorithm. \approach groups mined properties 
(and their associated signals) according to the first time they are violated. 
The elbow method implements a simple heuristic. 
Given a fixed error threshold, it starts by computing $k$-means clustering for $k=1$. 
The method increases the number of clusters as long as the sum of squared errors of the 
current clusters with respect to observed data is larger than the error threshold.
The suspicious signals from the same time clusters 
are then inductively associated to the Simulink blocks that contain them as well as to all their
block ancestors in the model hierarchy. 

Figure~\ref{fig:elbow} shows the diagram returned by the elbow method in our
case study, confirming that the violations are best clustered into $2$ groups.
The concrete clusters (not shown here) returned by the elbow method precisely
match the two groups we can intuitively entail from Table~\ref{tab:summary}.

\begin{table}[tb]
\begin{center}
        \begin{tabular}{c|l|l|c}
                Index & Signal Name & Block & $\tau (s)$ \\
                \hline
                $s_{252}$ & LI\_pos\_fail:1$\rightarrow$Switch:2 & Meas.  Left In. Act.  Pos. & 1.99 \\
                $s_{253}$ & Outlier/failure:1$\rightarrow$Switch:1 & Meas.  Left In. Act.  Pos. & 1.99 \\
                $s_{254}$ & Measured Position3:1$\rightarrow$Mux:3 & Meas.  Left In. Act.  Pos. & 1.99 \\
                $s_{255}$ & Measured Position2:1$\rightarrow$Mux:2 & Meas.  Left In. Act.  Pos. & 1.99 \\
                $s_{256}$ & Measured Position1:1$\rightarrow$Mux:1 & Meas.  Left In. Act.  Pos. & 1.99 \\
                $s_{55}$  & BusSelector:2$\rightarrow$Mux1:2 & Controller & 2.03 \\
                $s_{328}$ & In2:1$\rightarrow$Mux1:2 & L\_pos\_failures & 2.03 \\
                $s_{329}$ & In1:1$\rightarrow$Mux1:1 & L\_pos\_failures & 2.03 \\
                $s_{332}$ & Right Outer Pos. Mon.:2$\rightarrow$R\_pos\_failures:1 & Actuator Positions & 2.03 \\
                $s_{333}$ & Right Inner Pos. Mon.:2$\rightarrow$R\_pos\_failures:2 & Actuator Positions & 2.03 \\
                $s_{334}$ & Left Outer Pos. Mon.:2$\rightarrow$L\_pos\_failures:1 & Actuator Positions & 2.03 \\
                $s_{335}$ & Right Inner Pos. Mon.:3$\rightarrow$Goto3:1 & Actuator Positions & 2.03 \\
                $s_{338}$ & Left Outer Pos. Mon.:3$\rightarrow$Goto:1 & Actuator Positions & 2.03 \\
                $s_{341}$ & Left Inner Pos. Mon.:2$\rightarrow$L\_pos\_failures:2 & Actuator Positions & 2.03 \\
                $s_{272}$ & LO\_pos\_fail:1$\rightarrow$Switch:2 & Meas.  Left Out. Act.  Pos. & 3.99 \\
                $s_{273}$ & Outlier/failure:1$\rightarrow$Switch:1 & Meas.  Left Out. Act.  Pos. & 3.99 \\
                $s_{275}$ & Measured Position1:1$\rightarrow$Mux:1 & Meas.  Left Out. Act.  Pos. & 3.99 \\
                $s_{276}$ & Measured Position2:1$\rightarrow$Mux:2 & Meas.  Left Out. Act.  Pos. & 3.99 \\
                $s_{277}$ & Measured Position3:1$\rightarrow$Mux:3 & Meas.  Left Out. Act.  Pos. & 4.00 \\
        \end{tabular}

\caption{Internal signals that violate at least one learned invariant and Simulink blocks to which they belong. The column $\tau(s)$ denotes
the first time that each signal participates in an invariant violation.}
\label{tab:summary}
\end{center}
\vspace{-40pt}
\end{table}

\begin{figure}[t]
\centering
\includegraphics[width=\linewidth]{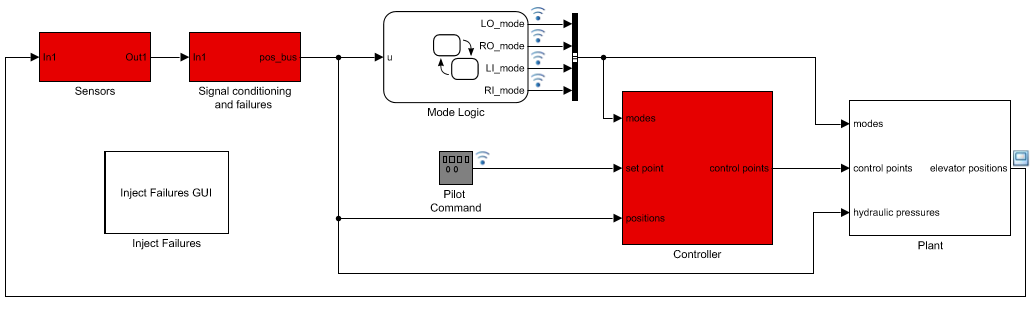}.
\\[\baselineskip]
\includegraphics[width=.45\linewidth]{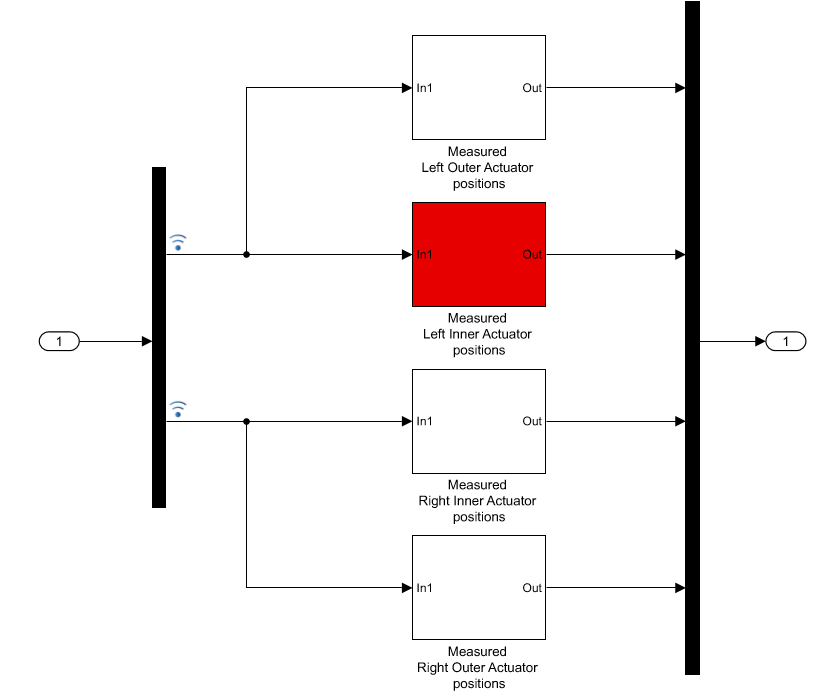}\quad\includegraphics[width=.45\linewidth]{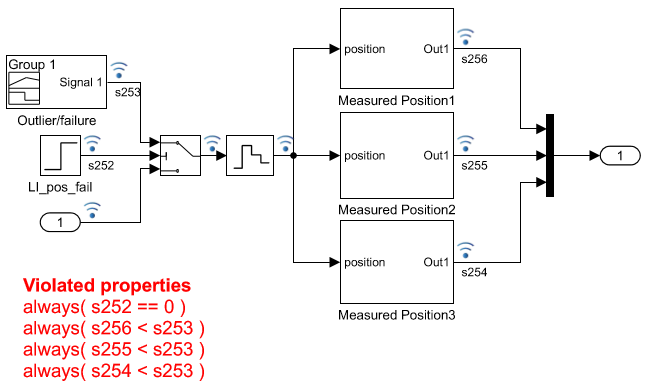}
\caption{Failure explanation as a sequence of snapshots - part of the first snapshot.}
\label{fig:visual}
\vspace{-10pt}
\end{figure}

Finally, \approach generates failure explanations that capture how the fault originated and propagated in space and time. In particular, the failure explanation is a sequence of snapshots of the system, one for each cluster of new property-violations. Each snapshot reports (i) the mean time as approximative time when the violations represented in the cluster occurred,
(ii) the model blocks that originate the violations reported in the cluster, (iii) the properties violated by the cluster, representing the reason why the cluster of anomalies exist, and (iv) the internal signals that participate to the violations of the properties associated with the cluster. Intuitively a snapshot represents a new relevant state of the system, and the sequence shows how the execution progresses from the violation of set of properties to the final violation of the specification. The engineer is supposed to exploit the sequence of snapshots to understand the failure, and the first snapshot to localize the root cause of the problem. 
Figure~\ref{fig:visual} shows the first snapshot of the failure explanation that \approach generated for the case study. 
We can see that the explanation of the failure at time $2$ involves the Sensors block, and propagates to Signal conditioning and failures and Controller blocks. By opening 
the Sensors block, we can immediately see that something is wrong with the sensor that measures the left inner position of the actuator. Going one level below, we can 
that the signal $s252$ coming out of the $\emph{LI\_pos\_fail}$ is suspicious -- indeed the fault was injected exactly in that block at time $2$. It is not a surprise that 
the malfunctioning of the sensor measuring the left inner position of the actuator affects the Signal conditioning and failures block (the block that detects if there is a 
sensor that fails) and the Controller block. However, at time $2$ the failure in one sensor does not affect yet the correctness of the overall system, hence the STL specification 
is not yet violated. The second snapshot (not shown here) generated by \approach reveals that the sensor measuring the left outer position of the actuator fails at 
time $4$. The redundancy mechanism is not able to cope with multiple sensor faults, hence anomalies become manifested in the observable behavior. 
From this sequence of snapshots, the engineer can conclude that the problem is in the failure of the two sensors - one measuring 
the left inner and the other measuring the left outer position of the actuator that stop functioning at times $2$ and $4$, respectively.

%% file: testing.pdf_t
\begin{picture}(0,0)%
\includegraphics{testing.pdf}%
\end{picture}%
\setlength{\unitlength}{3947sp}%
\begingroup\makeatletter\ifx\SetFigFont\undefined%
\gdef\SetFigFont#1#2#3#4#5{%
  \reset@font\fontsize{#1}{#2pt}%
  \fontfamily{#3}\fontseries{#4}\fontshape{#5}%
  \selectfont}%
\fi\endgroup%
\begin{picture}(8688,8424)(-311,-8473)
\put(3901,-3586){\makebox(0,0)[b]{\smash{{\SetFigFont{12}{14.4}{\rmdefault}{\mddefault}{\updefault}{\color[rgb]{0,0,0}$\psi_1$}%
}}}}
\put(3901,-4486){\makebox(0,0)[b]{\smash{{\SetFigFont{12}{14.4}{\rmdefault}{\mddefault}{\updefault}{\color[rgb]{0,0,0}$\psi_j$}%
}}}}
\put(1201,-6586){\makebox(0,0)[b]{\smash{{\SetFigFont{12}{14.4}{\rmdefault}{\mddefault}{\updefault}{\color[rgb]{0,0,0}$\psi_1$}%
}}}}
\put(1201,-7486){\makebox(0,0)[b]{\smash{{\SetFigFont{12}{14.4}{\rmdefault}{\mddefault}{\updefault}{\color[rgb]{0,0,0}$\psi_j$}%
}}}}
\put(3901,-6736){\makebox(0,0)[b]{\smash{{\SetFigFont{12}{14.4}{\rmdefault}{\mddefault}{\updefault}{\color[rgb]{0,0,0}P}%
}}}}
\put(3901,-8236){\makebox(0,0)[b]{\smash{{\SetFigFont{12}{14.4}{\rmdefault}{\mddefault}{\updefault}{\color[rgb]{0,0,0}their verdicts, and violation intervals}%
}}}}
\put(3901,-8011){\makebox(0,0)[b]{\smash{{\SetFigFont{12}{14.4}{\rmdefault}{\mddefault}{\updefault}{\color[rgb]{0,0,0}Internal signals in the fail trace}%
}}}}
\put(3901,-7636){\makebox(0,0)[b]{\smash{{\SetFigFont{12}{14.4}{\rmdefault}{\mddefault}{\updefault}{\color[rgb]{0,0,0}$F, I_k$}%
}}}}
\put(6901,-7711){\makebox(0,0)[b]{\smash{{\SetFigFont{12}{14.4}{\rmdefault}{\mddefault}{\updefault}{\color[rgb]{0,0,0}$B_p$}%
}}}}
\put( 76,-1561){\rotatebox{90.0}{\makebox(0,0)[b]{\smash{{\SetFigFont{20}{24.0}{\rmdefault}{\mddefault}{\updefault}{\color[rgb]{0,0,0}Testing}%
}}}}}
\put(1201,-1186){\makebox(0,0)[b]{\smash{{\SetFigFont{12}{14.4}{\rmdefault}{\mddefault}{\updefault}{\color[rgb]{0,0,0}$w^{1}_I$}%
}}}}
\put(1201,-2086){\makebox(0,0)[b]{\smash{{\SetFigFont{12}{14.4}{\rmdefault}{\mddefault}{\updefault}{\color[rgb]{0,0,0}$w^{n}_I$}%
}}}}
\put(1201,-1636){\makebox(0,0)[b]{\smash{{\SetFigFont{12}{14.4}{\rmdefault}{\mddefault}{\updefault}{\color[rgb]{0,0,0}$\cdots$}%
}}}}
\put(3901,-1186){\makebox(0,0)[b]{\smash{{\SetFigFont{12}{14.4}{\rmdefault}{\mddefault}{\updefault}{\color[rgb]{0,0,0}$w^{1}$}%
}}}}
\put(3901,-2086){\makebox(0,0)[b]{\smash{{\SetFigFont{12}{14.4}{\rmdefault}{\mddefault}{\updefault}{\color[rgb]{0,0,0}$w^{n}$}%
}}}}
\put( 76,-4261){\rotatebox{90.0}{\makebox(0,0)[b]{\smash{{\SetFigFont{20}{24.0}{\rmdefault}{\mddefault}{\updefault}{\color[rgb]{0,0,0}Mining}%
}}}}}
\put(5251,-736){\makebox(0,0)[b]{\smash{{\SetFigFont{12}{14.4}{\rmdefault}{\mddefault}{\updefault}{\color[rgb]{0,0,0}$\varphi$}%
}}}}
\put( 76,-7036){\rotatebox{90.0}{\makebox(0,0)[b]{\smash{{\SetFigFont{20}{24.0}{\rmdefault}{\mddefault}{\updefault}{\color[rgb]{0,0,0}Explaining}%
}}}}}
\put(3901,-1636){\makebox(0,0)[b]{\smash{{\SetFigFont{12}{14.4}{\rmdefault}{\mddefault}{\updefault}{\color[rgb]{0,0,0}$\cdots$}%
}}}}
\put(6601,-1636){\makebox(0,0)[b]{\smash{{\SetFigFont{12}{14.4}{\rmdefault}{\mddefault}{\updefault}{\color[rgb]{0,0,0}$\cdots$}%
}}}}
\put(6601,-1111){\makebox(0,0)[b]{\smash{{\SetFigFont{12}{14.4}{\rmdefault}{\mddefault}{\updefault}{\color[rgb]{0,0,0}$w^{1}$}%
}}}}
\put(6601,-2011){\makebox(0,0)[b]{\smash{{\SetFigFont{12}{14.4}{\rmdefault}{\mddefault}{\updefault}{\color[rgb]{0,0,0}$w^{n}$}%
}}}}
\put(6376,-6136){\makebox(0,0)[lb]{\smash{{\SetFigFont{12}{14.4}{\rmdefault}{\mddefault}{\updefault}{\color[rgb]{0,0,0}$t_1$}%
}}}}
\put(6376,-7336){\makebox(0,0)[lb]{\smash{{\SetFigFont{12}{14.4}{\rmdefault}{\mddefault}{\updefault}{\color[rgb]{0,0,0}$t_k$}%
}}}}
\put(1201,-4036){\makebox(0,0)[b]{\smash{{\SetFigFont{12}{14.4}{\rmdefault}{\mddefault}{\updefault}{\color[rgb]{0,0,0}$\cdots$}%
}}}}
\put(1201,-4486){\makebox(0,0)[b]{\smash{{\SetFigFont{12}{14.4}{\rmdefault}{\mddefault}{\updefault}{\color[rgb]{0,0,0}$w^{m}$}%
}}}}
\put(1201,-3586){\makebox(0,0)[b]{\smash{{\SetFigFont{12}{14.4}{\rmdefault}{\mddefault}{\updefault}{\color[rgb]{0,0,0}$w^{1}$}%
}}}}
\put(6676,-7036){\makebox(0,0)[b]{\smash{{\SetFigFont{12}{14.4}{\rmdefault}{\mddefault}{\updefault}{\color[rgb]{0,0,0}$\cdots$}%
}}}}
\put(2551,-6136){\makebox(0,0)[b]{\smash{{\SetFigFont{12}{14.4}{\rmdefault}{\mddefault}{\updefault}{\color[rgb]{0,0,0}$w^{m+1}$}%
}}}}
\put(6901,-6511){\makebox(0,0)[b]{\smash{{\SetFigFont{12}{14.4}{\rmdefault}{\mddefault}{\updefault}{\color[rgb]{0,0,0}$B_1$}%
}}}}
\put(3901,-4036){\makebox(0,0)[b]{\smash{{\SetFigFont{12}{14.4}{\rmdefault}{\mddefault}{\updefault}{\color[rgb]{0,0,0}$\cdots$}%
}}}}
\put(1201,-7036){\makebox(0,0)[b]{\smash{{\SetFigFont{12}{14.4}{\rmdefault}{\mddefault}{\updefault}{\color[rgb]{0,0,0}$\cdots$}%
}}}}
\put(7801,-6511){\makebox(0,0)[b]{\smash{{\SetFigFont{12}{14.4}{\rmdefault}{\mddefault}{\updefault}{\color[rgb]{0,0,0}$B_2$}%
}}}}
\put(3901,-7036){\makebox(0,0)[b]{\smash{{\SetFigFont{12}{14.4}{\rmdefault}{\mddefault}{\updefault}{\color[rgb]{0,0,0}$\cdots$}%
}}}}
\put(3901,-6511){\makebox(0,0)[b]{\smash{{\SetFigFont{12}{14.4}{\rmdefault}{\mddefault}{\updefault}{\color[rgb]{0,0,0}$w^{m+1}_1$}%
}}}}
\put(3901,-7411){\makebox(0,0)[b]{\smash{{\SetFigFont{12}{14.4}{\rmdefault}{\mddefault}{\updefault}{\color[rgb]{0,0,0}$w^{m+1}_k$}%
}}}}
\end{picture}%

%% file: experimental.tex
\section{Empirical Evaluation}
\label{sec:experimental}

We empirically evaluated our approach against three classes of faults: \emph{multiple hardware faults in fault-tolerant systems}, which is the case of multiple components that incrementally fail in a system designed to tolerate multiple malfunctioning units; \emph{incorrect lookup tables}, which is the case of lookup tables containing incorrect values; and \emph{erroneous guard conditions}, which is the case of imprecise conditions in the transitions that determine the state-based behavior of the system. Note that these classes of faults are highly heterogenous. In fact, their analysis requires a technique flexible enough to deal with multiple failure causes, but also with the internal structure of complex data structures and finally with state-based models.

We consider two different systems to introduce faults belonging to these three classes. We use the fault-tolerant aircraft elevator control system~\cite{mosterman-fdir} presented in Section~\ref{sec:caseStudy} to study the capability of our approach to identify failures caused by multiple overlapping faults. In particular, we study cases obtained by 
(1)  injecting a low pressure fault into two out of three hydraulic components (fault $h_{1}h_{2}$), and (2) inserting a fault in the left inner and left outer sensor 
position components (fault $\emph{lilo}$). 

We use the automatic transmission control system~\cite{HoxhaAF14} to study the other classes of faults. Automatic transmission control system is composed of $51$ variables, includes $4$ lookup tables of size between $4$ and $110$ and  two finite state machines running in parallel with $3$ and $4$ states, respectively, as well as $6$ transitions each. We used 
the $7$ STL specifications defined in~\cite{HoxhaAF14} to reveal failures in this system. We studied cases obtained obtained by (1) modifying a transition guard in the 
StateFlow chart (fault $\emph{guard}$), and (2) altering an entry in the look-up table Engine (fault $\emph{eng\_lt}$).

To study these faults, we considered two use scenarios. For the aicraft elevator control system, we executed $100$ test cases in which we systematically changed the 
amplitude and the frequency of the pilot command steps. These tests were executed on a non-faulty model. We then executed an additional test on the model to which 
we dynamically injected $h_{1}h_{2}$ and $\emph{lilo}$ faults. For the automatic transmission 
control system, we executed $100$ tests in which we systematically changed the step input of the throttle by varying the amplitude, the offset and the absolute time of the step. 
All the tests were executed on a faulty model. In both cases, we divided the failed tests from the passing tests. We used the data collected for the passing tests to infer models and the data obtained from the failing tests to generate failure explanations. 

We evaluated the output produced by our approach considering three main aspects: Scope Reduction, Cause Detection, Quality of the Analysis and Computation Time. Scope Reduction measures how well our approach narrows down the number of elements to be inspected to a small number of anomalous signals that require the attention of the engineer, in comparison to the set of variables involved in the failed execution. Cause detection indicates if the first cluster of anomalous values reported by our approach includes any property violation caused by the signal that is directly affected by the fault. Intuitively, it would be highly desirable that the first cluster of anomalies reported by our technique includes violations caused by the root cause of the failure. For instance, if a fault directly affects the values of the signal \texttt{Right Inner Pos.}, we expect these values to cause a violation of a property about this same signal. We qualitatively discuss the set of violated properties reported for the various faults and explain why they offer a comprehensive view about the problem that caused the failure.   
Finally, we analyze the computation time of \approach and its components and compare it to the simulation time of the model. 

To further confirm the effectiveness of our approach, we contacted $3$ engineers from (1) an automotive OEM with over 300.000 employees ($E1$), 
(2) a major modeling and simulation tool vendor with more than 3.000 employees ($E2$)
(3) an SME that develops tools for verification and testing of CPS models ($E3$). We asked them to evaluate the outcomes of our tool for a selection of faults (it was infeasible to ask them to inspect all the results we collected). In particular, we sent them the faulty program, an explanation of both the program and the fault, and the output generated by our tool\footnote{The report submitted to the engineers can be found in the Appendix.}, and we asked them to answer the following questions:  
\begin{enumerate}[label=Q\arabic*]
\item How helpful is the output to understand the cause(s) of the failure? (Very useful/Somewhat useful/Useless/Misleading)
\item Would you consider experimenting our tool with your projects? (Yes/May be/No)
\item Considering the sets of violations that have been reported, is there anything that should be removed from the output? (open question)
\item Is there anything more you would like to see in the output produced by our tool?  (open question)
\end{enumerate}

In the following, we report the results that we obtained for each of the analyzed aspects.

\subsection{Scope Reduction, Cause Detection and Qualitative Analysis}

Table~\ref{tab:reduction} shows the degree of reduction achieved for the analyzed faults. Column \emph{system} indicates the faulty application used in the evaluation. Column 
\emph{\# vars} indicates the size of the model in terms of the number of its variables. Column \emph{fault} indicates the specific fault analyzed. Column \emph{\# $\psi$} gives the number of learned invariants. Column \emph{\# suspicious vars} indicates the number of variables involved in the violated properties. Column \emph{fault detected} indicates whether 
the explanation included a variable associated to the output of the block in which the fault was injected.

\input{tableReduction}

We can see from Table~\ref{tab:reduction} that \approach successfully detected the exact origin of the fault in 3 out of 4 cases. 
In the case of the aircraft elevator control system, \approach clearly identifies the problem with the respective sensors (fault $\emph{lilo}$) 
and hydraulic components (fault $h_1 h_2$). The scope reduction amounts to $96\%$ and $90\%$ of the model signals for the $\emph{lilo}$ and 
the $h_1 h_2$ faults, respectively, allowing the engineer to focus on a small subset of the suspicious signals.

In the case of the automatic transmission control, \approach associates the misbehavior of the model to the Engine look-up table and points to its right entry. 
The scope reduction in this case is $90\%$.
On the other hand, \approach misses the exact origin of the $\emph{guard}$ fault and fails to point to the altered transition. This happens because the 
faulty guard alters only the {\em timing} but not the {\em qualitative} behavior of the state machine. Since Daikon is able to learn only invariant 
properties, \approach is not able to discriminate between passing and failing tests in that case. Nevertheless, \approach does associate the entire 
state machine to the anomalous behavior, since the observable signal that violates the STL specification is generated by the state machine.

\subsection{Computation Time}

Table~\ref{tab:time} summarizes computation time of \approach applied to the two case studies. We can make two main conclusions from these experimental 
results: (1) the overall computation time of \approach-specific activities is comparable to the overall simulation time and (2) property mining dominates by far 
the computation of the explanation. We finally report in the last row the translation of the Simulink simulation traces recorded in the Common Separated Values (csv) 
format to the specific input format that is used by Daikon. In our prototype implementation of \approach, we use an inefficient format translation that results in excessive times. 
We believe that investing an additional effort can result in improving the translation time by several orders of magnitude.

\input timing

\subsection{Evaluation by Professional Engineers}

We analyze in this section the feedback provided by engineers $E1-3$ to the questions $Q1-4$.

\begin{enumerate}[label=Q\arabic*]
\item $E1$ found \approach potentially very useful. $E2$ and $E3$ found \approach somewhat useful.
\item All engineers said that they would experiment with \approach.
\item None of the engineers found anything that should be removed from the tool outcome.
\item $E2$ and $E3$ wished to see better visual highlighting of suspicious signals. $E2$ wished to see the actual trace for each suspicious signal. 
$E2$ and $E3$ could not clearly understand the cause-effect relation from the tool outcome and wished a clearer presentation of cause-effects. 
\end{enumerate}

Apart from the direct responses to $Q1-4$, we received other useful information. All engineers shared appreciation for the visual presentation of outcomes, and 
especially the marking of suspicious Simulink blocks in red. $E1$ highlighted that real production models typically do not only contain Simulink and StateFlow blocks, but 
also SimEvent and SimScape blocks, Bus Objects, Model Reference, Variant Subsystems, etc., thus limiting the practical value of the current tool implementation. 

Overall, engineers confirmed that \approach can be a useful technology. At the same time, they offered valuable feedback to improve it, especially the presentation 
of the output produced by the tool.

%% file: tableReduction.tex
\begin{table}
\begin{center}
	\begin{tabular}{|l| c| c|| c| c| c| c|}
	\hline

	\textbf{system} & \textbf{\# vars} &  \textbf{fault} & \textbf{\# $\psi$} & \textbf{\# suspicious vars} & \textbf{fault detected} \\
\hline

	\multirow{2}{*}{aircraft} & \multirow{2}{*}{$426$} & $\emph{lilo}$ & \multirow{2}{*}{$96$} & $17$ & $\checkmark$ \\
	 &  &  $h_{1}h_{2}$ &  & $44$ & $\checkmark$     \\

\hline
	\multirow{2}{*}{transmission} & \multirow{2}{*}{$51$} & $\emph{guard}$ & \multirow{1}{*}{$41$} & $1$ &  \\
	& & $\emph{eng\_lt}$ & 39 & $4$ & $\checkmark$ \\
	
\hline
	\end{tabular}

\caption{Scope reduction and cause detection.}
\label{tab:reduction}
\end{center}
\vspace{-40pt}
\end{table}

%% file: timing.tex
\begin{table}
	\begin{center}
		\begin{tabular}{|l|c|c|}
			\hline
			&  aircraft & transmission \\
			\hline
			\# tests & $151$ & $100$ \\
			\# samples per test & $1001$ & $751$ \\
			\hline
			& \multicolumn{2}{|c|}{time (s)} \\
			\hline
			Simulation & $654$ & $35$ \\
			\hline
			Instrumentation & $1$ & $0.7$ \\
			Mining & $501$ & $52$ \\
			Monitoring properties & $0.7$ & $0.6$ \\
			Analysis & $1.5$ & $1.6$ \\
			\hline
			File format translation & $2063$ & $150$ \\
			\hline
		\end{tabular}
	\end{center}
	\label{tab:time}
	\caption{\approach computation time.}
	\vspace{-40pt}
\end{table}

%% file: related.tex

\section{Related Work}
\label{sec:related}

The analysis of software failures has been addressed with two main classes of related approaches: fault localization and failure explanation techniques.

\smallskip

\emph{Fault localization} techniques aim at identifying the location of the faults that caused one or more observed failures (an extensive survey can be found in~\cite{WongGLAW16}). A popular example is \emph{spectrum-based fault-localization} (SFL)~\cite{4344104}, an efficient statistical technique that, by measuring the code coverage in the failed and successful tests, can rank the program components (e.g., the statements) that are most likely responsible for a fault.

SFL has been recently employed to localize faults in Simulink/Stateflow CPS models~\cite{BartocciFMN18,DeshmukhJMP18,LiuLNBB16,LiuLNBB16b,LiuLNB17}, showing similar accuracy as in the application to software systems~\cite{LiuLNBB16b}. The explanatory power of this approach is however limited, because it generates neither information that can help the engineers understanding if a selected code location is really faulty nor information about how a fault propagated across components resulting on an actual failure.  
Furthermore, SFL is agnostic to the nature of the oracle requiring to know only whether the system passes or not a specific test case.  This prevents the exploitation of any additional information concerning why and when the oracle decides that the test 
is not conformed with respect to the desired behavior. In Bartocci et al.~\cite{BartocciFMN18} the authors try to overcome 
this limitation by assuming that the oracle is a monitor generated from an STL specification. This approach  allows the use of the trace diagnostic method proposed in Ferr\`{e}re et al.~\cite{FerrereMalerNickovic15}  to obtain more information (e.g., the time interval when the 
cause of violation first occurs) about the failed tests improving the fault-localization. Although this additional knowledge can improve the confidence on the localization, still little is known about the root cause of the problem and its impact on the runtime behavior of the CPS model.

\approach complements and improves SFL techniques generating information that helps engineers identifying the cause of failures, understanding how faults resulted in chains of anomalous events that eventually led to the observed failures, and producing a corpus of information well-suited to support engineers in their debugging tasks, as confirmed by the subjects who responded to our questionnaire.

\smallskip

\emph{Failure explanation techniques} analyze software failures in the attempt of producing information about failures and their causes. For instance, a few approaches combined mining and dynamic analysis in the context of component-based and object-oriented applications to reveal~\cite{DBLP:conf/issta/PastoreMHFSSM14} and explain failures~\cite{BefroueiWW16,BCT,AVA}. These approaches are not however straightforwardly applicable to CPS models, since they exploit the discrete nature of component-based and object-oriented applications that is radically different from the data-flow oriented nature of CPS models, which include mixed-analog signals, hybrid (continuous and discrete) components, and a complex dynamics.

\approach originally addresses failure explanation in the context of CPS models. The closest work to \approach is probably Hynger~\cite{JohnsonBD15,NguyenHBDJ18}, which exploits invariant generation to detect specification mismatches, that is, a mismatch between an actual and an inferred specification, in Simulink models. Specification mismatches can indicate the presence of problems in the models. Differently from Hynger, \approach does not compare specifications but exploits inferred properties to identify anomalous behaviors in observed failures. Moreover, \approach exploits correlation and clustering techniques to maintain the output compact, and to generate a sequence of snapshots that helps comprehensively defining the story of the failure. Our results show that this output can be the basis for cost-effective debugging. 

%% file: conclusions.tex
\section{Future Work and Conclusions}
\label{sec:conclusions}

We have presented \approach, an automatic approach for explaining failures 
in Simulink models.  Our approach combines testing, specification mining 
and failure analysis to provide a concise explanation consisting 
of time-ordered sequence of model snapshots that show the variable exhibiting 
anomalous behavior and their propagation in the model.  We evaluated 
the effectiveness \approach on two models, involving two use scenarios and 
several classes of faults.

We believe that this paper opens several research directions.  In this work, we 
only considered mining of invariant specifications.  However, we have observed 
that invariant properties are not sufficient to explain timing issues, hence we plan 
to experiment in future work with mining of {\em real-time temporal} specifications. 
In particular, we will study the trade-off between the finer characterization of the 
model that temporal specification mining can provide and its computational cost. 
We also plan to study systematic ways to explain failures in presence of heterogeneous 
components.  In this paper, we consider the setting in which we have multiple 
passing tests, but we only use a single fail test to explain the failure.  We will study 
whether the presence of multiple failing tests can be used to improve the explanations. 
In this work, we have performed manual fault injection and our focus was on studying the 
effectiveness of \approach on providing meaningful failure explanations for different 
use scenarios and classes of faults. 
We plan in the future to develop automatic fault injection and perform systematic 
experiments for evaluating how often \approach is able to find the root cause.